\documentstyle[12pt,epsf,epsfig]{article}

\font\fivesy=cmsy5

\newcommand{\p}{\bot}
\newcommand{\pa}{\scriptscriptstyle \|}
\newcommand{\paa}{\hbox{\textfont2=\fivesy$\|$}}
\newcommand{\dd}{\partial}

\newcommand{\de}{\delta}
\newcommand{\De}{\Delta}
\newcommand{\om}{\omega}
\newcommand{\Om}{\Omega}
\newcommand{\e}{\varepsilon}
\newcommand{\f}{\varphi}
\newcommand{\ls}{\left(}
\newcommand{\lks}{\left[}
\newcommand{\rs}{\right)}
\newcommand{\rks}{\right]}
\newcommand{\g}{\gamma}
\newcommand{\al}{\alpha}
\newcommand{\be}{\beta}
\newcommand{\ta}{\tau}
\newcommand{\n}{\nu}
\newcommand{\m}{\mu}
\newcommand{\s}{\sigma}
\newcommand{\La}{\Lambda}
\newcommand{\la}{\lambda}

\newcommand{\et}{\eta}

\newcommand{\ps}{\psi}
\newcommand{\str}[1]{\mathrel{\mathop{\longrightarrow}\limits_{#1}}}

\newcounter{form}
\newcommand{\dis}[1]{$$\displaylines{#1}$$}
\newcommand{\disn}[2]{$$\displaylines{\refstepcounter{form}
            \label{#1} \hfill #2}$$}
\newcommand{\no}{\hfill \phantom{(\theform)}\cr \hfill}
\newcommand{\nom}{\hfill (\theform) \cr}

\newcounter{punkt}
\makeatletter
\renewcommand{\section}{\@startsection{section}{1}{0pt}%
            {3.5ex plus 1ex minus .2ex}{2.3ex plus .2ex}{\bf}}

\long\def\@makecaption#1#2{%
   \vskip 10\p@
   \setbox\@tempboxa\hbox{#1. #2}%
   \ifdim \wd\@tempboxa >\hsize
       #1. #2\par
     \else
       \hbox to\hsize{\hfil\box\@tempboxa\hfil}%
   \fi}

\makeatother
\newcommand{\sect}[2]{\protect\refstepcounter{punkt}\protect\label{#1}
            \section*{$\protect\vphantom{a}$\hfill
            \arabic{punkt}.\hskip 2mm #2 \hfill $\protect\vphantom{a}$}}
\newcommand{\st}{\hfill $\protect\vphantom{a}$\protect\\
            $\protect\vphantom{a}$\hfill}

\begin{document}

\title{Constructing the light-front QCD Hamiltonian}

\author{S.~A.~Paston\thanks{E-mail: paston@pobox.spbu.ru},
V.~A.~Franke\thanks{E-mail: franke@pobox.spbu.ru},
E.~V.~Prokhvatilov\thanks{E-mail: Evgeni.Prokhvat@pobox.spbu.ru}\\
St.-Petersburg State University, Russia}

\date{}

\maketitle
\begin{abstract}
We propose  the  light-front  Lagrangian  and  the  corresponding
Hamiltonian that produce a theory  perturbatively  equivalent  to
the  conventional  QCD  in  the  Lorentz  coordinates  after  the
regularization is removed. The regularization used is nonstandard
and breaks the gauge invariance. But after the regularization  is
removed, this invariance is restored by  the  introduction  of  a
finite number of counterterms with coefficients dependent on  the
regularization parameters.
\end{abstract}

\vfill
\noindent
In this corrected version the counterterms are given
in more exact form.

\vskip 20mm
\noindent
Original version is
published in Theoretical and Mathematical Physics,\\
Vol.~120, No.~3, pp.~1164-1181, 1999;

\noindent
translated from Teoreticheskaya i Maternaticheskaya Fizika,\\
Vol.~120, No.~3, pp.~417-437, September, 1999.

\newpage

\sect{vved}{Introduction}

The search for ways to solve problems in the quantum field theory
with a large  coupling  constant,  has  long  been  significant.
Calculations on space-time lattices are now often used  for  this
purpose in  QCD.  Essential  results  have  been  thus  obtained.
Nevertheless, these calculations are very laborious  and  have  a
low accuracy; in addition, it is generally difficult to  estimate
the calculation error theoretically. Therefore, it is interesting
to seek other possible approaches to this problem.  Even  limited
progress in this direction  would  allow  comparing  the  results
obtained by different methods.

Long  before  the  advent  of  QCD,  studying  the   pion-nucleon
interaction  had  been  attempted  by  solving  the   Schrodinger
equation in the Lorentz coordinate system in the framework of the
quantum theory of  pion  and  nucleon  fields.  The  states  were
described by the method previously found by V.~A.~Fock \cite{fok}
in terms of vectors in the space that now bears his name. In this
case, the mathematical vacuum of the Fock  space  coincided  with
the  free  theory  vacuum.  This  approach,  now  known  as   the
Tamm-Dancoff method \cite{tam,dan}, did not lead to success.  The
primary reason was the complexity of the physical  vacuum  state,
which did not coincide with the mathematical vacuum.  Without  a
description  of  the  physical  vacuum,  it  was  impossible   to
investigate any other states. If ultraviolet  (UV)  and  infrared
(IR) cutoffs are introduced to make  the  number  of  degrees  of
freedom finite, it would be possible, in principle, to  represent
the physical vacuum as a vector of the Fock space with  the  free
theory vacuum. However, such a  representation  proves  extremely
complicated because it is necessary to  provide  the  translation
invariance of the physical vacuum and  to  satisfy  the  "cluster
decomposition property" for vacuum expectation values. For  these
reasons, the Schrodinger equation  in  the  Lorentz  coordinates,
where the evolution occurs in the conventional  time,  is  hardly
applicable to any calculations in the quantum field theory with a
large coupling constant.

As early as 1949, Dirac  suggested  a  method  for  avoiding  the
difficulties related to the description of  the  physical  vacuum
state \cite{dir}. He proposed using the light-front coordinates
$x^\pm=(x^0\pm x^3)/\sqrt{2}$, $x^1$, $x^2$,
where $x^0$, $x^1$, $x^2$, $x^3$ are  the  Lorentz
coordinates, and treating $x^+$ as time. In this approach, a theory
is canonically quantized on the  surface  $x^+=const$,  and  the
generator $P_+$ of the shift along the $x^+$ axis plays the role  of
the Hamiltonian $H$. In addition, the generator of the shift  along
the $x^-$ axis, i.e., the momentum operator $P_-$, does not  shift  the
surface $x^+=const$, where the quantization is performed,  and  is
kinematic, according to the Dirac terminology (in contrast to the
dynamic generator $P_+$). Therefore, the structure of  the  operator
$P_-$ in a theory with interaction  is  the  same  as  in  a  theory
without interaction, i.e., the operator $P_-$ is always quadratic in
the creation and annihilation operators ${a_n}^+(p_-,p_{\p})$
and $a_n(p_-,p_{\p})$ and, as a rule, has the form (after normal ordering)
\dis{
P_-=\int d^2p_{\p}\int\limits_0^{\infty} dp_-\; p_-
\sum_n {a_n}^+(p_-,p_{\p})a_n(p_-,p_{\p}),
}
where $p_{\p}=(p_1,p_2)$ and the index $n$ enumerates  the  species
of the creation and  annihilation  operators.  According  to  the
spectral  condition,  the  operator  $P_-$  is  positive   definite;
therefore, the integration over $p_-$
in the given formula is performed only from $0$ to $\infty$.
The operator $P_-$ vanishes on the physical vacuum $\Om$,
i.e., $P_-|\Om\rangle=0$, whence it follows that
$a_n(p_-,p_{\p})|\Om\rangle=0$.

For this reason, the physical vacuum $|\Om\rangle$ can  be  taken
as the mathematical vacuum of the Fock space  generated  by  the
operators ${a_n}^+$. No  question  of  describing  the  physical  vacuum
structure arises. The spectrum of the bound states can  be  found
by solving the Schrodinger equation
 \dis{
P_+|\Psi\rangle =p'_+|\Psi\rangle
}
under the conditions
 \dis{
P_-|\Psi\rangle=p'_-|\Psi\rangle,\quad
P_{\p}|\Psi\rangle=0,
}
where $p'_+$, $p'_-$ are numbers and the mass squared is defined by
 \dis{
m^2=2p'_+p'_-.
}
Here, $|\Psi\rangle$ is a  vector  in  the  just  mentioned  Fock
space. It is easy to satisfy the conditions
$P_-|\Psi\rangle=p'_-|\Psi\rangle$  and
$P_{\p}|\Psi\rangle=0$, where $p'_-$
is chosen arbitrarily. The problem  is  to
solve the Schrodinger  equation.  Such  an  approach  is  usually
called the "light-front Hamiltonian" approach. Obviously, it  can
also be used to calculate the scattering matrix.

The  scheme  described  faces  considerable  problems   due    to
divergences that arise and  the  lack  of  the  explicit  Lorentz
invariance.  But  the  possibility  of  avoiding    the    direct
description  of  the  physical  vacuum  structure  is   such    a
considerable advantage that this approach  continues  to  attract
attention. Interest in it has increased with the advent  of  QCD.
This paper is devoted to developing this approach.

We  note  that  the  light-front  coordinates  and  the   similar
"infinitely large-momentum system"  are  used  in  quantum  field
theory not only in the  framework  of  the  Hamiltonian  approach
based on directly solving the Schrodinger equation. Many  results
have been obtained by studying the limiting case  of  fast-moving
reference frames in the framework of  the  explicitly  Lorentz-invariant.
theory of the scattering  matrix  or  of  the  Green's
functions \cite{kva1,kva2,kva3,kva4,karm}.
But we consider only the Hamiltonian approach in
this paper.

For the theory with the given Lorentz-invariant  initial  action,
constructing the light-front Hamiltonian $P_+$  proves  a  difficult
problem. The primary reason is the specific divergences  at  zero
values of tin' momentum $p_-$ of virtual quanta. In particular,  the
invariant volume element on the hyperboloid $p_{\m}p^{\m}=m^2$ has  the
form $d^2p_{\p} dp_-/p_-$ and contains $p_-$ in
the denominator. The situation
becomes more complicated in gauge theories.  Even  in  the  first
papers on this problem \cite{tomb,kash}, it became clear  that  canonical
quantization on the surface $x^+=const$ could be performed only in
the gauge $A_-=0$ or in similar gauges (for example, $\dd_- A_-=0$).
The reason is that second-class constraints arise in  the  theory
and solving them, in particular, requires inverting the covariant
derivative $D_-=\dd_-+igA_-$. But in the gauge $A_-=0$, the  Feynman
propagator has  an  extra  term  $p_-$  in  the  denominator,  which
strengthens  the  singularities  at  $p_-=0$  (at  least,  in  the
perturbation theory).

As a consequence, a special  regularization  is  required,  which
consists in "cutting out"  the  neighborhood  of  $p_-=0$  in  the
momentum space one way or another and which results  in  breaking
the Lorentz invariance (until  the  regularization  is  removed).
This is unavoidable in the liglit-front Hamiltonian approach.  In
principle, we can preserve the gauge invariance  if,  instead  of
"cutting  out"  the  neighborhood  of $p_-=0$,  we  restrict  the
space-time with respect to the coordinate $x^-$ ($-L\le x^-\le L$)  and
impose periodic boundary
conditions in $x^-$ on all fields \cite{nov}. In this  case,  the
spectrum of the momentum  $P_-$  becomes  discrete,  and  the  "zero
modes" of the fields $A$, i.e., the Fourier modes corresponding  to
$p_-=0$,  are  explicitly  singled  out.  To  preserve  the  gauge
invariance, these zero modes must be taken into account.  It  was
shown in \cite{nov} that secondary second-class  constraints  arise  in
the theory, from which we  must  find  the  above-mentioned  zero
modes as functions of the other modes and substitute them in  the
Hamiltonian. These constraints arc so  complicated  that  solving
them proves impossible. Therefore, from the  very  beginning,  we
must discard the zero modes in the Lagrangian, which results in
breaking the gauge invariance. Therefore, the  breakdown  of  the
gauge invariance by a regularization is unavoidable in any  case.
In what follows, to regularize  the  theory,  we  "cut  out"  the
neighborhood  of  $p_-=0$.  In  addition,  the  conventional  UV
regularization is, of course, necessary.

It follows from the above discussion that  the  formal  canonical
quantization  on  the  hypersurface  $x^+=const$  can  produce  a
Hamiltonian corresponding to a theory that is not  equivalent  to
the initial  Lorentz-invariant one,  even  in  the  limit  of
removing the regularization. As a rule, the  equivalence  can  be
provided only by adding nonconventional counterterms
\cite{tmf,bur1,bur2} to the light-front Hamiltonian.

In the last few years, a number of papers have appeared in  which
the  approximate  regularized  light-front  QCD  Hamiltonian   was
directly constructed \cite{vils,perr}. Using  the  renormalization  group
theory, the form of the Hamiltonian was  adjusted  based  on  the
requirement that the result be weakly dependent on the UV cutoff,
with the relation to the  conventional  Lorentz-invariant  theory
not  being  traced  in  detail.  Methods  for  simplifying    the
Hamiltonian  and  solving  the  Schrodinger  equation  were  also
proposed.  The  technique  described  in  these  papers  is    of
considerable interest and continues to develop.  But  it  remains
unclear to what  extent  the  light-front  theory  thus  obtained
corresponds to the conventional Lorentz-invariant QCD.

In this connection, we  meet  the  problem  of  constructing  the
light-front Hamiltonian such that it produces a theory equivalent
to  the  conventional  Lorentz-invariant  one  in  the  limit  of
removing the regularization. As a necessary condition, we  should
first  provide  this  equivalence  in  the  framework   of    the
perturbation  theory.  Then,  in  particular,  we  can  use   the
technique  described  in \cite{vils,perr} to  simplify  the   obtained
Hamiltonian and the subsequent nonperturbative calculations based
on the Schrodinger equation. In doing so, we see the relation  to
the conventional Lorentz-invariant theory.

The problem of constructing the counterterms for the  light-front
Hamiltonian, which provide the equivalence of this approach to
the Lorentz-invariant one, was investigated in \cite{tmf}  in  the
framework of the perturbation theory in the coupling constant. In
that paper, the authors  proposed  a  method  for  comparing  two
perturbation series for the Green's functions  constructed  using
the  light-front  Hamiltonian  for  one  and  the    conventional
Lorentz-invariant approach for the other. In addition to  the  UV
regularization, they regularized the  singularities  at  $p_-=0$
using the cutoff $|p_-|\ge \e >0$,  i.e.,  eliminating  the  Fourier
components with
$|p_-|<\e$ for every field in the theory. It  was  revealed  that
for the required equivalence for the nongauge field theories  (in
particular, for the Yukawa model), it was only necessary to add a
few counterterms to the canonical  light-front  Hamiltonian.  But
for gauge theories (both Abelian and non-Abelian) under the given
regularization, it proved  necessary  to  introduce  an  infinite
number of counterterms in  the  Hamiltonian.  This  situation  is
closely  related  to  the  gauge  condition $A_-=0$,  which   is
unavoidable in the  canonical  light-front  quantization.  It  is
known \cite{bas1,bas2}  that  to  correctly  construct  the  perturbation
theory in this gauge, it is necessary  to  use  the  gauge  field
propagator in the form proposed by Mandelstam and Leibbrandt \cite{man,lei}
  \disn{v.0}{
\frac{-i\de^{ab}}{k^2+i0}\ls
g_{\m\n}-\frac{k_\m n_\n+k_\n n_\m}{2k_+k_-+i0}2\frac{k_+}{n_+}\rs=
\frac{-i\de^{ab}}{k^2+i0}\ls
g_{\m\n}-\frac{k_\m n_\n+k_\n n_\m}{2(kn^*)(kn)/(nn^*)+i0}2
\frac{(kn^*)}{(nn^*)}\rs
\nom}
where $\mu,\nu =+, -, 1, 2$; $k_\pm =(k_0\pm k_3)/\sqrt{2}$;
$n_{-,1,2}=0$, $n^*_{+,1,2}=0$.
The additional pole in $k_-$ in this expression is cut  out  with
the regularization $|k_-|\ge \e$. The  distortion  arising  in  this
case does not disappear in the limit $\e \to 0$. An infinite  number
of counterterms is required to compensate this distortion.

The aim of this paper is to overcome this difficulty  and  obtain
the required Hamiltonian with a finite number of counterterms. In
such a case, we must change the regularization. For this purpose,
we propose shifting the pole with respect to $k_-$ in  expression
(\ref{v.0}) from the point $k_-=0$ by changing the Lagrangian such that a
small regularizing mass parameter $\m^2$ is added  to  the  quantity
$2k_+k_-$. In this case, the distortions caused by cutting out  the
interval $|k_-|<\e$ are not so large as in the preceding case,  and
a finite number of  counterterms  is  sufficient  to  obtain  the
correct Hamiltonian in the limit of removing the regularization
($\e \to 0$ and then $\m \to 0$ together with $\La\to \infty$,
where  $\La$  is  a  UV
regularization parameter). We choose the UV  regularization  such
that, after removing the IR regularization (i.e., after setting
$\e=0$ and $\m=0$) at the intermediate stages,  we  obtain  the
Lorentz-invariant Lagrangian regularized in the UV region.  This
increases the number of "ghosts," but  the  number  of  necessary
counterterms would otherwise increase sharply. We  note  that  in
the gauge $A_-=0$, the one-particle irreducible vertex parts  with
the  upper  index  "$-$"  do  not  contribute  to  the   Green's
functions. In what follows, we prove  the  coincidence  of  every
order of the perturbation theory in the  limit  of  removing  the
regularization only for the Green's functions  and  not  for  the
vertex parts with the index "$-$". This is sufficient  because  the
masses of the bound states are determined by  the  poles  of  the
Green's functions.

Under the given regularization, the Hamiltonian acts on the space
with an indefinite metric, which prevents using the  conventional
variational principle to solve the Schrodinger equation under the
conditions of preserving the regularization. Nevertheless,  there
exist different  variational  methods  that  allow  solving  this
equation.

The chosen regularization breaks the local gauge  invariance  but
preserves the global  $SU(3)$  invariance.  For  this  reason,  the
number of necessary counterterms, being  finite,  is  essentially
larger than the conventional one. We show that there is a way  to
choose the  counterterms  such  that  the  obtained  light-front
theory  is  perturbatively    equivalent    to    the    initial
Lorentz-invariant one in the limit of removing the regularization.

We achieve this goal by starting with the Lagrangian
 \disn{v.1}{
L=L_0+L_I
\nom}
 \disn{v.2}{
L_0=-\frac{1}{4}\sum_{j=0,1}(-1)^jf_j^{a,\m\n}
\ls 1+\frac{\dd_{\pa}^2}{\La^2_j}-
\frac{\dd_\p^2}{\La^2}\rs f_{j,\m\n}^a+\no
+\sum_{l=0}^3\frac{1}{v_l}\bar\ps_l\ls i\g^\m\dd_\m-M_l\rs \ps_l
\nom}
 \disn{v.3}{
L_I=c_0\dd_\mu A^a_\n\dd^\mu A^{a,\n}+
c_{01}\dd_\mu A^a_\n\, N^{\m\al}\,\dd_\al A^{a,\n}+
c_1\dd_\mu A^{a,\mu}\,\dd_\n A^{a,\n}+
c_{11}\, N^{\m\al}\,\dd_\mu A^a_\al\,\dd_\n A^{a,\n}+\no+
c_{12}\, N^{\m\al}\,\dd_\mu A^a_\al
\, N^{\n\be}\,\dd_\n A^a_\be+
c_2 A^a_\mu A^{a,\mu}+
c_3 f^{abc} A^a_\mu A^b_\n \dd^\mu A^{c,\n}+
c_{31} f^{abc} A^a_\mu A^b_\n
\, N^{\al\m}\,\dd_\al A^{c,\n}+\no+
 A^a_\mu A^b_\n A^c_\g A^d_\de
\bigl(c_4 f^{abe}f^{cde} g^{\mu\g}g^{\n \de}+
\de^{ab}\de^{cd}\ls c_5 g^{\mu\g}g^{\n \de}+
c_6 g^{\mu\n}g^{\g\de}\bigr)\rs+\no
+c_7 \bar\ps \g^\m i\dd_\m\ps+
c_{71}\bar\ps \g_\m\, N^{\m\n}\,i\dd_\n\ps+
c_{72}\bar\ps \g_\m\, N^{\n\m}\,i\dd_\n\ps-
c_{8} \bar\ps\ps+\no+
c_{9}A^a_\m \bar\ps \g^\m \frac{\la^a}{2}\ps+
c_{91}A^a_\al\,N^{\m\al}\,\bar\ps\g_\m\frac{\la^a}{2}\ps,
\nom}
where
 \disn{v.4}{
f^a_{j,\m\n}=\dd_\m A^a_{j,\n}-\dd_\n A^a_{j,\m},\qquad
\frac{1}{\La_j^2}=
\cases{1/\La^2,
&$j=0$,\cr 1/\La^2+1/\m^2, &$j=1$,\cr}
\nom}\vskip -3mm
 \disn{v.4a}{
v_0=1,\quad \sum_{l=0}^3v_l=0,\quad \sum_{l=0}^3v_lM_l=0,\quad
\sum_{l=0}^3v_lM_l^2=0,
\nom}\vskip -3mm
 \disn{v.4b}{
A^a_\m=\sum_{j=0,1}A^a_{j,\m},\quad \ps=\sum_{l=0}^3\ps_l.
\nom}
Here
 \disn{isp7}{
\dd_{\pa}^2=2\dd_+\dd_-,\qquad \dd_\p^2=\dd_1^2+\dd_2^2,\qquad
N^{\al\be}=\frac{n^\al n^{*\be}}{(nn^*)},
\nom}
$\g^\m$ are the Dirac matrices, $\la^a$ are the matrices of the fundamental
representation of the gauge $SU(3)$ group, and
 \disn{v.7}{
{\rm Tr}\la^a=0,\quad
\la^a\la^b=if^{abc}\la^c+d^{abc}\la^c+\frac{2}{3}\de^{ab},\quad
{\rm Tr}(\la^a\la^b)=2\de^{ab}.
\nom}
We assume that the fields $A^a_{j,\m}$, are restricted by the condition
 \disn{v.5}{
A^a_{j,-}=0.
\nom}
In addition, we introduce the cutoff in the momenta $k_-$ and $k_{\p}$
 \disn{v.6}{
\e\le |k_-|\le V, \qquad v^2\le k_{\p}^2\le V^2
\nom}
for all fields from the very beginning. This cutoff  as  well  as
condition (\ref{v.5}) excludes certain degrees of freedom directly in the
Lagrangian; such a procedure does not lead to new constraints  in
the canonical formalism, because no variation with respect to the
excluded degrees of freedom is performed.

The quantities $\e$, $\La$, $\m$, $v$, $V$  and  $M_1$-$M_3$  are
regularization  parameters.  The  coefficients  $c_i$   are
renormalization  constants,  i.e.,  they  are  functions  of  the
regularization parameters and  are  expansions  in  the  coupling
constant $g$. These expansions begin with $g$ (with  the  coefficient
one) for $c_3$ and $c_{9}$  and  with  $g^2$  and  higher  for  the  others.
Expression (\ref{v.3}) contains the conventional QCD interaction and  the
additional terms necessary for the  renormalizability  under  the
assumption that the Lorentz invariance and the  global,  but  not
local, gauge invariance are preserved.

In Sec.~\ref{dok}, we  prove  that  the  light-front  Hamiltonian
corresponds to the given Lagrangian.  In  the  framework  of  the
perturbation theory with respect to the coupling constant  $g$  and
with a certain dependence of the renormalization constants on the
regularization parameters, this Hamiltonian produces the Green's
functions of the fields $A^a_{0,\m}$, $\bar\ps_0$, and $\ps_0$
coinciding  with  the
Green's functions of the conventional QCD  (renormalized  in  the
Lorentz coordinates using dimensional  regularization)  in  every
order  with  respect  to  $g$  in  the  limit  of  removing    the
regularization (according to the special prescription).

\sect{gam}{The light-front Hamiltonian}

We develop the canonical light-front  formalism  for  the  theory
with Lagrangian (\ref{v.1})-(\ref{v.3})
defined  on  the  fields  satisfying  the
condition $A^a_{j,-}=0$  and  subject  to  cutting  out  the  momenta
according to formula (\ref{v.6}).
We use the following representation for the bispinors $\ps_l$ and the
matrices $\g^\m$:
 \disn{1.7d}{
\ps=\ls
\begin{array}{c}
\ps_{+}\\
\ps_{-}
\end{array}
\rs,\qquad
\g^0=\ls
\begin{array}{cc}
0&-i\\
i&0
\end{array}
\rs,\quad
\g^3=\ls
\begin{array}{cc}
0&i\\
i&0
\end{array}
\rs,\quad
\g^k=\ls
\begin{array}{cc}
-i\s_k&0\\
0&i\s_k
\end{array}
\rs.
\nom}
Here and in what follows, the indices $k,l$ ranges over $1,2$.

Let us rewrite the expression for the $L$ in a form more convenient for
the canonical light-front  formalism development using instead
of the $A_{j,+}^a$ new variables
 \disn{isp8}{
\f^a_j=\dd_- A^a_{j,+}-\dd_k A^a_{j,k}.
\nom}
For example, for the $L_I$ we get:
 \disn{isp1}{
L_I=c'_{01}A^a_k\dd_-\dd_+A^a_k-c_0A^a_k\dd^2_\p A^a_k+
c_1\f^a\f^a+c_{11}\ls \f^a+\dd_k A^a_k\rs \f^a+\no+
c_{12}\ls \f^a+\dd_k A^a_k\rs\ls \f^a+\dd_l A^a_l\rs
-c_2A^a_k A^a_k+
c_3 f^{abc} A^a_\mu A^b_\n \dd^\mu A^{c,\n}+
c_{31} f^{abc} A^a_\mu A^b_\n
\, N^{\al\m}\,\dd_\al A^{c,\n}+\no+
 A^a_\mu A^b_\n A^c_\g A^d_\de
\bigl(c_4 f^{abe}f^{cde} g^{\mu\g}g^{\n \de}+
\de^{ab}\de^{cd}\ls c_5 g^{\mu\g}g^{\n \de}+
c_6 g^{\mu\n}g^{\g\de}\bigr)\rs+\no+
c'_{71}i\sqrt{2} \ps_+^+\dd_+\ps_++
c'_{72}i\sqrt{2} \ps_-^+\dd_-\ps_-+
c_7\ls\ps_+^+ i\hat\dd_\p\ps_-+h.c\rs-
c_8\ls i\ps_-^+\ps_++h.c\rs+\no+
c'_{91}\sqrt{2}A_+^a\ps_+^+\frac{\la^a}{2}\ps_++
c_9\ls\ps_+^+\hat A_\p\ps_-+h.c.\rs,
\nom}
where
 \disn{isp2}{
\f^a=\sum_{j=0,1}\f^a_j,\qquad
\hat A_\p\equiv A_k^a\ls\frac{\la^a}{2}\s_k\rs,\qquad
\hat\dd_\p\equiv\ls\s_k\dd_k\rs,\no
c'_{01}=2c_0+c_{01},\qquad
c'_{71}=c_7+c_{71},\qquad
c'_{72}=c_7+c_{72},\qquad
c'_{91}=c_9+c_{91}.
\nom}
We transform  the  part  of  the
Lagrangian that contains the derivatives $\dd_+$
of the fermion fields
$\ps$, presenting it as a diagonalized bilinear form:
 \disn{1.7a}{
\sum_{l=0}^3\frac{1}{v_l}\ps_{l+}^+ \dd_+\ps_{l+}
+c'_{71} \ps_+^+ \dd_+\ps_+=\!
\sum_{l=0}^3B_{ll'}\ps_{l+}^+ \dd_+\ps_{l'+}=\!
\sum_{l=0}^3 \frac{1}{w_l}\ps'^+_{l+}\dd_+\ps'_{l+},
\nom}
where
 \disn{1.7c}{
\ps_{l+}=\sum_{l'=0}^3 U_{ll'}\ps'_{l'+},
\nom}
$1/w_l$ are the eigenvalues of the matrix $B_{ll'}$,
and $U_{ll'}$ is the matrix of the transformation diagonalizing $B_{ll'}$.

Consider the  part
of the Dirac equation that is a  result  of  the  variation  with
respect to $\ps_{l-}^+$:
 \disn{isp3}{
\frac{1}{v_l}\ls \sqrt{2}\dd_-\ps_{l-}+\hat\dd_\p\ps_{l+}-M_l\ps_{l+}\rs
+c'_{72}\sqrt{2}\dd_-\ps_-+c_7\hat\dd_\p\ps_+-c_8\ps_+-ic_9\hat A_\p\ps_+=0.
\nom}
This equation does not contain derivatives with  respect  to
$x^+$. It is a constraint that can be used to express $\ps_{l-}$
in terms of other variables.
To do this we sum the equation (\ref{isp3})
over index $l$ with the weight $v_l$.
Using the equations (\ref{v.4a}) we obtain
 \disn{isp4}{
\sqrt{2}\dd_-\ps_-=-\sum_{l'=0}^3\ls \hat\dd_\p -M_{l'}\rs\ps_{l'+}.
\nom}
By the substitution of this expression to the equation (\ref{isp3})
one can obtain easily from the latter
 \disn{1.8}{
\ps_{l-}=\frac{1}{\sqrt{2}}\dd_-^{-1}Y_l,
\nom}
where
 \disn{1.8a}{
Y_l=-\ls\hat\dd_\p-M_l\rs\ps_{l+}
+v_l\lks c'_{72}\sum_{l'=0}^3\ls\hat\dd_\p-M_{l'}\rs\ps_{l'+}-
c_7\hat\dd_\p\ps_++c_8\ps_++ic_9\hat A_\p\ps_+\rks.
\nom}
Let us remark that such a constraint can be resolved
in this way not only in $A_-=0$ gauge but in any gauge,
if Pauli-Villars fermions are present (if no Pauli-Villars
fermions we need to invert the operator
$D_-=\dd_--igA_-$, and therefore to use the $A_-=0$ gauge).

In turn, the $\ps_+$ is expressed in terms of the $\ps'_{l+}$ by (\ref{v.4b})
and (\ref{1.7c}). After
substituting the expression (\ref{1.8}) into Lagrangian
the latter depends only on the variables $A^a_{j,\m}$ and $\ps'_{l+}$.

We eliminate the derivatives $\dd^2_+A^a_{j,k}$
in the Lagrangian
by integrating by parts and find the momenta conjugate
to the $A^a_{j,k}$ (by corresponding variation of the $L$
at fixed $\f^a_j$):
 \disn{1.6}{
\Pi^a_{j,k}=\frac{\dd L}{\dd(\dd_+A^a_{j,k})}=\frac{(-1)^j}{\La_j^2}
\Bigl[ 4\dd_-^2\dd_+A^a_{j,k}-\dd_\p^2\dd_-A_{j,k}^a+
\La_j^2\ls 1-\frac{\dd_\p^2}{\La^2}\rs\dd_- A^a_{j,k}\Bigr]
-c'_{01}\dd_- A^a_k.
\nom}
If we similarly define the momenta conjugate to  $\f^a_j$,  we
obtain  the  second-class  constraint.   Using    the    Legendre
transformation with respect to the variables  $\dd_+ A^a_{j,k}$
and  $\Pi^a_{j,k}$,
which  does  not  affect  the  variable  $\dd_+\f^a_j$,
we  pass  to  the
first-order  Lagrangian. This Lagrangian depends on the
derivatives $\dd_+\ps'_l$,  $\dd_+\f^a_j$,  and  $\dd_+A^a_{j,k}$
only  linearly  (the
dependence on $\dd_+\ps'_l$ and $\dd_+\f^a_j$
is linear from the beginning). Using
the Fourier transformation-type formulas, we then pass to the new
variables, playing the  role  of  the  creation  and  annihilation
operators in order that the following  conditions  be  satisfied.
First, the part of the  Lagrangian  containing  derivatives  with
respect  to  $x^+$  must  have  a  standard  canonical  form.   This
automatically  solves  the  above-mentioned  constraint  for  the
variable  $\f^a_j$.  Second,  the  positive-definite    free-part must be
diagonal in the corresponding Fock space. The  following  changes
of variables meet these conditions:
 \disn{1.12}{
A^a_{j,k}(x)=\frac{1}{(2\pi)^{3/2}}
\!\!\int\limits_{\e\le p_-\le V}\!\!dp_-
\!\!\int\limits_{v^2\le p_\p^2 \le V^2}\!\!d^2p_\p
\sum_{r=0,1}\frac{a^a_{jr,k}(p)e^{-ipx}+h.c.}{\sqrt{2\om_j}},\no
\Pi^a_{j,k}(x)=\frac{-i}{(2\pi)^{3/2}}
\!\!\int\limits_{\e\le p_-\le V}\!\!dp_-
\!\!\int\limits_{v^2\le p_\p^2 \le V^2}\!\!d^2p_\p
\sum_{r=0,1}(-1)^r\sqrt{\frac{\om_j}{2}}\lks a_{jr,k}(p)e^{-ipx}-h.c.\rks,\no
\f_{j}^a(x)=\frac{i\La_j}{(2\pi)^{3/2}}
\!\!\!\int\limits_{\e\le p_-\le V}\!\!\!dp_-
\!\!\!\int\limits_{v^2\le p_\p^2 \le V^2}\!\!\!d^2p_\p
\frac{a_{j}^a(p)e^{-ipx}-h.c.}{\sqrt{2p_-}},\no
\ps'^i_{l+,s}(x)=
\frac{2^{-1/4}}{(2\pi)^{3/2}}
\!\!\!\int\limits_{\e\le p_-\le V}\!\!\! dp_-
\!\!\!\int\limits_{v^2\le p_\p^2 \le V^2}\!\!\! d^2p_\p
\ls b^i_{l,s}(p)e^{-ipx}+{d^i_{l,s}}^+(p)e^{ipx}\rs,
\nom}
where $\om_j=p_-\left|1-j \frac{p_\p^2}{\m^2}\right|=p_-\ls\frac{p_\p^2}{\m^2}-1\rs^j$,
$O(p)\equiv O(p_-,p_\p)$, $px\equiv p_-x^-+p_kx^k$,
the index $s$ enumerates the spinor
components ($s=1,2$), and $i$  is  the  index  of  the  fundamental
representation of the color group. All creation and  annihilation
operators are defined for $p_-\hskip -3pt\ge\hskip -3pt\e$.
We also assume that $v>\m$ in
these formulas. The resulting Lagrangian has the form
 \disn{isp5}{
L=i\!\!\int\limits_{\e\le p_-\le V}\!\!dp_-
\!\!\int\limits_{v^2\le p_\p^2 \le V^2}\!\!d^2p_\p\;
\Biggl\{-\sum_{j,a}(-1)^j {a^a_{j}}^+(p)\dd_+ a^a_{j}(p)+\no
+\sum_{j,r,a,k}(-1)^r {a^a_{jr,k}}^+(p)\dd_+ a^a_{jr,k}(p)+
\sum_{l,i}\frac{1}{w_l}\ls
{b^i_l}^+(p)\dd_+ b^i_l(p)+{d^i_l}^+(p)\dd_+ d^i_l(p)\rs\Biggr\}-H,
\nom}
where $H=P_+$ is a Hamiltonian.

Accordingly, the (anti)commutation relations have the form
 \disn{1.15}{
[a^a_{jr,k}(p),a^{a'}_{j'r',k}(p')]=
[a^a_{jr,k}(p),a^{a'}_{j'}(p')]=
[a^a_{jr,k}(p),{a^{a'}_{j'}}^+(p')]=0,\no
[a^a_{jr,k}(p),{a^a_{j'r',k'}}^+(p')]=(-1)^r\de^{aa'}\de_{jj'}\de_{rr'}
\de_{kk'}\de(p_--p'_-)\de^{(2)}(p_\p-p'_\p),\no
[a^a_{j}(p),{a^a_{j'}}^+(p')]=-(-1)^j\de^{aa'}\de_{jj'}
\de(p_--p'_-)\de^{(2)}(p_\p-p'_\p),\no
\{b^i_{l,s}(p),{b^{i'}_{l',s'}}^+(p')\}=
\{d^i_{l,s}(p),{d^{i'}_{l',s'}}^+(p')\}=\hskip 40mm\no\hskip 40mm
=w_l\de^{ii'}\de_{ll'}\de_{ss'}
\de(p_--p'_-)\de^{(2)}(p_\p-p'_\p).
\nom}
The negative signs  in  the  right-hand  side  of  the
(anti)commutation relations correspond to the degrees of  freedom
carrying an indefinite metric in the space  of  states,  and  the
corresponding operators are "ghosts".

In contrast to the conventional canonical light-front  formalism,
this formalism contains no constraints that halve the  number  of
the Fourier components of the  fields  $A^a_{j,k}$.  To  preserve  the
positivity of the momentum $p_-$ everywhere, we are forced to double
the number of the creation and annihilation operators ${a^a_{jr,k}}^+$  and
$a^a_{jr,k}$ by introducing the index $r=0,1$. The  first-order  part  of
the Lagrangian, which contains no  derivatives  with  respect  to
$x^+$, coincides with the Hamiltonian $H=P_+$ up to a sign. It has
the form
 \disn{1.11}{
H=\int dx^-\int d^2x^\p\sum_{j=0,1}\Biggl\{
\frac{(-1)^j\La_j^2}{8}\biggl[
(-1)^j\ls\dd_-^{-1}\Pi^a_{j,k}+c'_{01}A^a_k\rs+
\frac{\dd_\p^2}{\La_j^2}A^a_{j,k}-
\Bigl( 1-\frac{\dd_\p^2}{\La^2}\Bigr) A^a_{j,k}\biggr]^2+\no+
\frac{(-1)^{j+1}}{2}A_{j,k}^a\dd_\p^2\Bigl( 1-\frac{\dd_\p^2}{\La^2}\Bigr)A^a_{j,k}
+\frac{(-1)^{j+1}}{2}\f_j^a\Bigl( 1-\frac{\dd_\p^2}{\La^2}\Bigr)\f_j^a\Biggr\}+
c_0A_k\dd_\p^2A_k+\no+
\frac{i}{\sqrt{2}}\sum_{l=0}^3\frac{1}{v_l}\ps_{l+}^+\ls\dd_\p^2-M_l^2\rs
\dd_-^{-1}\ps_{l+}-
\frac{i}{\sqrt{2}}c'_{72}\sum_{l,l'=0}^3\ps_{l+}^+\ls\hat\dd_\p+M_l\rs
\dd_-^{-1}\ls\hat\dd_\p-M_{l'}\rs\ps_{l'+}+\nom+
\ls\frac{i}{\sqrt{2}}c_{7}\ps_+^+\hat\dd_\p\dd_-^{-1}\sum_{l=0}^3
\ls\hat\dd_\p-M_l\rs\ps_{l+}+h.c.\rs+
\ls\frac{i}{\sqrt{2}}c_{8}\ps_+^+\dd_-^{-1}\sum_{l=0}^3
\ls\hat\dd_\p-M_l\rs\ps_{l+}+h.c.\rs+\no+
\sqrt{2}c'_{91}\ls\f^a+\dd_kA_k^a\rs\dd_-^{-1}\ls\ps_+^+\frac{\la^a}{2}\ps_+\rs+
\ls\frac{1}{\sqrt{2}}c_9\ps_+^+\hat A_\p\dd_-^{-1}\sum_{l=0}^3\ls\hat\dd_\p-M_l\rs
\ps_{l+}+h.c.\rs
-L'_I.
\no}
Here, $L'_I$ denotes expression (\ref{isp1}) with the  terms  with
the coefficients $c_0$, $c'_{01}$, $c_7$, $c'_{71}$, $c'_{72}$, $c_8$,
$c_9$, $c'_{91}$ omitted,
$\ps_{l+}$ is expressed in terms of $\ps'_{l+}$
by  formulas  (\ref{v.4b})
and (\ref{1.7c}), and the quantities
$A^a_{j,k}$, $\Pi^a_{j,k}$, $\f^a_{j}$, $\ps'_{l+}$
and  $\ps'^+_{l+}$
are expressed in terms of the creation and annihilation operators
by formulas (\ref{1.12}).

The operator $P_-$ has the form
 \disn{1.20}{
P_-=\int\limits_{\e\le p_-\le V}\!dp_-\!
\int\limits_{v^2\le p_\p^2 \le V^2}\!d^2p_\p\; p_-\Biggl\{\sum_{j=0,1}(-1)^j
\biggl[-a^a_{j}{}^+(p)a^a_{j}(p)+\no+
\sum_{r=0,1}(-1)^{j+r} {a^a_{jr,k}}^+(p)a^a_{jr,k}(p)\biggr]+\no+
\sum_{l=0}^3 \frac{1}{w_l}
\ls {b^i_l}^+(p)b^i_l(p)+{d^i_l}^+(p)d^i_l(p)\rs\Biggr\}.
\nom}
This operator is positive definite in the  Fock  space  with  the
vacuum defined by
 \disn{1.19}{
a^a_{j}|0\rangle=a^a_{jr,k}|0\rangle=b^i_l|0\rangle
=d^i_l|0\rangle\equiv 0.
\nom}

In the framework of the perturbation  theory,  in  the  limit  of
removing the regularization, all "ghosts" are switched off in the
sense that the Green's functions  of  this  theory  tend  to  the
Green's functions of the correct theory, which has  no  "ghosts".
This  gives  us  hope  that  in  the  limit  of   removing    the
regularization, the unitarity condition  also  holds  beyond  the
scope of the perturbation theory.

\sect{dok}{Comparison of the light-front and Lorentz-invariant\st
perturbation theories}

By the result of the perturbation theory, we mean the set of  the
Green's functions for the fields $A^a_{0,\m}$, $\bar\ps_0$,
and  $\ps_0$  constructed
perturbatively in the coupling constant $g$. We regard  the  fields
$A^a_{1,\m}$, $\bar\ps_l$ and $\ps_l$ for $l=1,2,3$  as  auxiliary.
We  show  that
Hamiltonian (\ref{1.11}) with a certain $\La$
dependence of the  coefficients
$c_i$ produces a perturbation theory coinciding in the limit
of removing the regularization with the renormalized perturbation
theory obtained from the conventional QCD Lagrangian in the gauge
$A_-=0$  using  the  Mandelstam-Leibbrandt   prescription    and
dimensional regularization (the {\it conventional} perturbation  theory
in what follows). (See \cite{bas1,bas2} for the  renormalization  of  the
conventional perturbation
theory.) The regularization is removed as follows:
first, $V\to \infty$, then $\e\to 0$, then $\La\to \infty$;
$M_1$, $M_2$,  $M_3$,  $\m$ and $v$ are assumed to be functions
of $\La$ such that $M_1,M_2,M_3\to\infty$, $\m\to 0$ and $v\to  0$
as $\La\to \infty$. The latter two functions  must  satisfy  more
exact  restrictions:  $v>\m$  (this  condition  was   used    in
constructing  the  Hamiltonian)  and  $\m\La,v\La\str{\La\to\infty}0$
(these restrictions are obtained below).

First, the  noncovariant  perturbation  theory  produced  by  the
Hamiltonian can  be  obtained  from  the Feynman  perturbation
theory constructed based on the Lagrangian corresponding  to  the
given Hamiltonian by resumming diagrams in every order and  using
the following integration rule: in calculating the  diagrams,  we
first integrate over $k_+$ (the momentum component conjugate to  the
light-front time $x^+$) and then over  the  other
components  \cite{har,lay}.
Therefore, it is sufficient to prove that  the  perturbation
theory obtained from Lagrangian (\ref{v.1})-(\ref{v.3})
and supplemented  by  the
given integration rule and  the  limiting  transitions  coincides
with the conventional perturbation theory.

We represent the free part of the Lagrangian, Eq.~(\ref{v.2}), in a  form
convenient for the perturbation theory analysis,
 \disn{2.1}{
L_0=\frac{1}{2}\sum_{j=0,1}(-1)^jA^a_{j,\m}
\ls 1+\frac{\dd_{\pa}^2}{\La^2_j}-\frac{\dd_\p^2}{\La^2}\rs
\ls g^{\m\n}\dd^2-\dd^\m\dd^\n\rs
A^a_{j,\n}+\no
+\sum_{l=0}^3\frac{1}{v_l}\bar\ps_l\ls i\g^\m\dd_\m-M_l\rs \ps_l,
\nom}
and we  separate  the  part  corresponding  to  the  conventional
interaction from (\ref{v.3}),
 \disn{2.2}{
L^{conv}_I=gf^{abc}A^a_\m A^b_\n \dd^\m A^{c,\n}-
\frac{g^2}{4}f^{abe}f^{cde}A^a_\m A^b_\n A^{c,\m}A^{d,\n}-
g A^a_\m \bar\ps \g^\m \frac{\la^a}{2}\ps.
\nom}
We assume that the remaining part of (\ref{v.3}) (where all  coefficients
are the expansions in $g$ starting from  the  terms  of  order  $g^2$)
consists of the renormalization counterterms, i.e., it eliminates
the divergences arising in the perturbation theory as  $\La\to\infty$.
The notation used and the additional conditions adopted are given
by formulas (\ref{v.4})-(\ref{v.6}) and in the text following them.

The propagators of the fields $A^a_{j,\m}$ and $\ps_l$ in the momentum
space are
 \disn{2.4}{
\Delta^{ab}_{j,\rho\n}=
\frac{-i\de^{ab}}{k^2+i0}\ls g_{\rho\n}-\frac{k_\rho n_\n+k_\n n_\rho}{k_-}\rs
\frac{(-1)^j}{1-\frac{k_{\paa}^2}{\La_j^2}+\frac{k_\p^2}{\La^2}-i0},
\nom}
 \disn{2.5}{
\Delta^{\ps}_l=iv_l\frac{\g^\m k_\m+M_l}{k^2-M_l^2+i0},
\nom}
where $n_+=1$  and $n_-,n_\p=0$.

Because the fields $A^a_{j,\m}$ and $\ps_l$ always
enter interaction (\ref{v.3})
in terms of the sums $A^a_\m$ and $\ps$ the sums
of the propagators
 \dis{
\Delta^{ab}_{\rho\n}\equiv\sum_j\Delta^{ab}_{j,\rho\n}
\qquad {\rm and} \qquad
\Delta^{\ps}\equiv\sum_l\Delta^{\ps}_l
}
always enter the diagrams as in the Pauli-Villars regularization.
After  all  diagrams  are  presented  in  terms  of  the  summary
propagators, we can take the limit  $V\to \infty$  for  an  arbitrary
diagram (i.e., remove the
restrictions $|k_-|\le  V$ and  $k_\p^2\le  V^2$) because,  in  view  of
conditions (\ref{v.4a}), the propagator $\Delta^{\ps}$
decreases  sufficiently  fast
(the sufficiently fast decrease of propagators (\ref{2.4})  is  provided
by the finiteness of $\La$) and the integrals converge.

The summary propagator $\Delta^{ab}_{\rho\n}$ is
 \disn{2.6}{
\Delta^{ab}_{\rho\n}=
\frac{-i\de^{ab}}{k^2+i0}\ls\frac{k^2_{\pa}}{k^2_{\pa}-\hat\m^2+i0}
g_{\rho\n}-\frac{k_\rho n_\n+k_\n n_\rho}{k^2_{\pa}-\hat\m^2+i0}2k_+\rs
R,
\nom}
where
 \disn{2.7}{
\hat \m^2=\m^2\frac{\La^2+k_\p^2}{\La^2+\m^2},\quad
R=\frac{1}{\ls 1-\frac{k^2}{\La^2}-i0\rs \ls 1+\frac{\m^2}{\La^2}\rs}.
\nom}
After removing the cutoff ($\La\to \infty$), we have  $\hat\m\to  0$,
and the propagator takes the conventional form. In terms  of  the
propagators $\Delta^{ab}_{\rho\n}$ and $\Delta^{\ps}$
and the vertices from (\ref{2.2}),  the  set  of
Feynman diagrams is the same as in the conventional  perturbation
theory.

It is easy to see that as long as $\La$  is finite, there  are  no  UV
divergences in the  perturbation  theory  constructed  using  the
Lagrangian under consideration. Using this fact, as well  as  the
condition $k_\p^2\ge  v^2$ (see~(\ref{v.6})),
we  can  apply  the  formalism
presented in \cite{tmf} to the perturbation theory and show  that  for
the majority of diagrams, after the limit $\e\to 0$ is  taken,  the
result  of  their  light-front  calculation  (where  we    first
integrate over $k_+$ and then over the other  components  according
to the rules providing the correspondence with  the  noncovariant
perturbation theory as explained above) coincides with the result
of calculating the same diagrams in the Lorentz coordinates.  The
possible discrepancy that arises in some cases can be compensated
by redefining the coefficient $c_2$ in the Lagrangian. This is shown
in Appendix~1. Therefore, it is sufficient to prove that  in  the
limit  $\La\to  \infty$,  the  perturbation  theory  with  the  summary
propagators $\Delta^{ab}_{\rho\n}$ and $\Delta^{\ps}$,
with interaction (\ref{v.3}), and  with  the
restrictions $\e\le |k_-|\le V$ and $k_\p^2\le  V^2$
removed coincides with  the
conventional perturbation theory. We emphasize that  we  can  now
perform all calculations in the Lorentz coordinates;  therefore,
we can make  the  Wick  rotation  and  pass  to  calculating  the
diagrams in the Euclidean space (the location  of  the  poles  of
propagators (\ref{2.6}) allows this).

Propagator (\ref{2.6}) differs from the propagator of  the  conventional
perturbation theory, first, in the  factor  $R$  providing  the  UV
regularization, second, in the quantity $\hat \m$, which vanishes  as
$\La\to\infty$ and, third, in the condition $k_\p^2\ge v^2$,
where $v\to 0$  as $\La\to\infty$.

We now analyze the behavior of an arbitrary Feynman diagram as $\m\to 0$
and $v\to 0$ (for finite  $\La$).  In  this  case,  essential  IR
divergences (essential in the  sense  that  they  arise  for  any
values of external momenta and not just for special  values)  can
occur.  If  such  a  divergence  does  not appear,  then    in
investigating the limit $\La\to 0$ for an arbitrary diagram, we  can
at once set $\hat\m=0$ and $v=0$ in its integrand. In this case,  the
error in the integrand contains the factor $\hat  \m^2$. The UV divergence
of the initial diagram is not worse  than  quadratic;  therefore,
after separating the factor $\hat  \m^2$,
where $\hat\m^2\sim\m^2(1+k_\p^2/\La^2)$,  the
divergence is not worse than logarithmic, and the  condition  for
an error decrease is $\m^2\ln\La\str{\La\to\infty}0$.
This consideration does
not take an increase in the IR divergence  after  separating  the
factor $\hat\m^2$ into account. Its  power  increases  by  two.  Because
there was no divergence before, this power becomes  not,  greater
than one, i.e., integrating the IR divergence gives (in  view  of
the factor $\hat\m^2$) the order  $\hat\m$.
Integrating  over  the  remaining
variables produces  an  UV  divergence  not  worse  than  linear.
Consequently, the condition for
the error decrease is $\m\La\str{\La\to\infty}0$.
Similar considerations give the condition $v\La\str{\La\to\infty}0$.

We analyze the occurrence of essential IR divergences as $\m,v\to  0$
in Appendix~2.
It turns out that such divergences arise only in
one case -- for the diagram terms which contain the following factor
 \disn{2.8}{
\De_{+\rho} G^{\rho\n}\De_{\n\al},
\nom}
where $G^{\rho\n}(k)$ is an arbitrary one-particle irreducible
two-point subdiagram, and $\De$ are propagators.
In which connection the divergence takes place only when index $\m$
in formula (\ref{2.8}) takes the values $1,2$, and the divergence
the whole of diagram at that is logarithmic:  $(\ln\m)^N$,
where $N$ is not larger then the number of factors of form (\ref{2.8}).

It is clear why such a divergence does not produce  any  problems
in the conventional perturbation theory, when $G^{\rho\n}(k)=G^{\rho\n}_{dim}(k)$
is calculated gauge invariantly with the aid of dimensional regularization.
In the expression (\ref{2.8}) at $\rho="-"$ the first propagator turns to zero,
and, besides, $g_{++}=g_{+\p}=0$, hence, one should consider in the propagator
only the item containing the sum $(k_+n_\rho+k_\rho n_+)$. The first term of this
sum does not give a divergence because $n_1=n_2=0$, and the second term gives
the factor $k_\rho G^{\rho\n}(k)$, which is equal to zero because  of  the
Ward  identities (analogous to ones adduced in work \cite{bas3}) which are
the consequences of exact maintenance of gauge invariance.
From the given reasoning it is clear that  breaking
the  gauge  invariance    without    the
simultaneous regularization of the essential IR divergences makes
the  perturbation  theory  senseless. In    our    case,    this
regularization is provided by introducing the quantity $\m$.

By induction on the loop number, we prove  that  with  a  certain
choice of the coefficients of the counterterms in Lagrangian (\ref{v.3})
in every order, the value of every Feynman diagram tends  to  its
conventional value calculated using dimensional regularization as
$\La\to\infty$. It is clear that in the one-loop  order,  there  are  no
subdiagrams; therefore, there are no  essential  IR  divergences.
Then on considering a diagram containing  factor  (\ref{2.8}),  for  the
lower order $G^{\rho\n}(k)$ subdiagram, we have
 \disn{2.9}{
G^{\rho\n}(k)-G^{\rho\n}_{dim}(k)=O\ls\frac{1}{\La}\rs
\nom}
at $\rho,\n\ne "-"$
(just  this  estimate  is  obtained   in    Appendix~4).
Because the value $G^{\rho\n}(k)$ at $\rho="-"$ or $\n="-"$
does not give a contribution to the expression (\ref{2.8})
then taking into account (\ref{2.9}) we can maintain that accurate to
$O\ls\frac{1}{\La}\rs$ the expression (\ref{2.8}) coincides with
 \disn{2.10}{
\De_{+\rho} G^{\rho\n}_{dim}\De_{\n\al},
\nom}
where, as it was already said, the divergence is absent.

Therefore if $(\ln\m)^N/\La\str{\La\to\infty}0$ then in the limit
$\La\to\infty$ the diagrams containing the factor of form (\ref{2.8})
will not differ from their values calculated in the conventional
perturbation theory under dimensional regularization.
More exactly a possible difference is due to
the diagram divergence, but it is polynomial and  is  compensated
by the counterterms  of  the  same  form  that  arise  under  the
renormalization. Therefore, it is now sufficient to prove that in
the limit $\La\to\infty$,  the  Euclidean  perturbation  theory  with
propagator (\ref{2.6}), where we set $\m=0$, with
the  propagator $\Delta^{\ps}$,
with  interaction  (\ref{v.3}),  and  with  restrictions (\ref{v.6}) removed
coincides with the conventional perturbation theory.

Propagator  (\ref{2.6}) with $\m=0$ after the transition to the
Euclidean space can be written down as
 \disn{isp6}{
\Delta^{ab}_{\rho\n}=
\frac{\de^{ab}}{k^2}\ls
\de_{\rho\n}-\frac{k_\rho n_\n+k_\n n_\rho}{(kn)}\rs
\frac{1}{1+\frac{k^2}{\La^2}}.
\nom}
This expression is Lorentz-invariant
if we assume that the vector $n_\n$
(complex  in  the  Euclidean  space,
such that $in_0-n_3=0$, $n_{1,2}=0$)  to  be    properly
transformed under the Lorentz transformation. It is
interesting that there is no distinguished vector other  than $n_\n$
in the Euclidean space, whereas in  the  pseudo-Euclidean  space,
the  Mandelstam-Leibbrandt   prescription    distinguishes    the
additional surface $k_+=0$.
However the vector  $n_\n$ is complex, and there is new
fixed vector, namely, the complex conjugated to $n_\n$.
It is seen that it coincides  (up to a factor) with Euclidean
continuation of the vector $n^*_\n$, which is defined below
the equation (\ref{v.0}) and picks out the surface $k_+ = 0$
in pseudoeuclidean space.

Interaction  Lagrangian (\ref{2.2}) is  also
Lorentz-invariant;
therefore, the counterterms that must be added to the  Lagrangian
under the renormalization in  every  order  of  the  perturbation
theory are Lorentz-invariant.
It is shown in the Appendix 4 that
the values which it is necessary to add to the diagrams
in order to
in the limit $\La\to\infty$ make them
finite and coincident
with their values calculated using dimensional regularization
are polynomials with respect to external momenta with the coefficients
containing factors $N_{\al\be}$ (see the definition of this value in
formula (\ref{isp7})).
We should take into account that in the counterterms
the vector $n_\n$ cannot  be contracted with the field
$A_{\n}$, because
we consider the Feynman diagrams for the  Green's  functions,  whose
external lines cannot carry the upper index "$-$" as it gives zero
after convolution with the propagators. It is  evident  that  the
counterterms are globally gauge  invariant  because  the  initial
Lagrangian has this property. We now analyze the possible form of
the counterterms.

There exist logarithmically divergent diagrams  with  four  gluon
external lines. In Appendix~3, we show that the structure  of  an
arbitrary diagram of this type with respect to the labels of  the
gauge group can
only have the form $f^{abe}f^{cde}$ or $\de^{ab}\de^{cd}$.
Taking the  Lorentz  invariance
into account,  we  conclude  that the  general  form  of  the
corresponding counterterms is exhausted by  the  terms  with  the
coefficients $c_4$, $c_5$, and  $c_6$  in  Lagrangian  (\ref{v.3})
and  we  can
therefore replace the addition of the counterterms by  a  certain
choice of these coefficients.

There also exist divergent diagrams  with  three  gluon  external
lines. In general, they can diverge linearly;  however,  because
of the Lorentz invariance, the divergent part  must  contain  the
factor $k^\m$, and the divergence is really logarithmic. We show  in
Appendix~3 that the structure of an  arbitrary  diagram  of  this
type with respect to the labels of the gauge group can only  have
the form $f^{abc}$. For the general  form  of  the  divergence,  this
gives the terms with  the  coefficients  $c_3$ and $c_{31}$
in  (\ref{v.3})
(note, that the existence of the latter takes account of the volume
$N_{\al\be}$ appearance in the counterterms),  and  we  can
therefore replace the  addition  of  counterterms  by  a  certain
choice  of  these  coefficients.  In    addition,    there    exist
logarithmically divergent diagrams with two fermion and one gluon
external lines. It is  evident  that  the  general  form  of  the
divergence is defined by the terms with the coefficients $c_{9}$ and
$c_{91}$ in  (\ref{v.3})
and we can therefore replace the addition of  counterterms  by  a
certain choice of these coefficients.

Next, there exist divergent  diagrams  with  two  gluon  external
lines. They diverge quadratically, and the general  form  of  the
divergence is given by the term $c_2$ in (\ref{v.3}). But after  subtracting
the quadratic divergence, a linear divergence can remain. Because
of the Lorentz invariance, it is  really  absent,  and  only  the
logarithmic divergence exists. It is  evident  that  the  general
form of this divergence is given by terms with  the  coefficients
$c_{0}$, $c_{01}$, $c_{1}$, $c_{11}$, $c_{12}$
in (\ref{v.3}) and we can therefore  replace  the  addition  of
counterterms by a certain choice of these coefficients and coefficient
$c_2$. There also exist divergent diagrams with two fermion external
lines. They  diverge  linearly,  and  the  general  form  of  the
divergence is given by the term with the coefficient $c_{8}$
in  (\ref{v.3}).
But  after  subtracting  the  linear  divergence,  a  logarithmic
divergence can remain. It is evident that  the  general  form  of
this divergence is given by the term with the coefficients
$c_{7}$, $c_{71}$, $c_{72}$ in
(\ref{v.3}) and we can therefore replace the addition of counterterms  by
a choice of these coefficients and coefficient $c_{8}$.

We conclude that the perturbation theory  with  propagator  (\ref{isp6}),
with the  propagator $\Delta^{\ps}$ and with  interaction
(\ref{v.3}) (we must compare  such  a
perturbation theory with the conventional one) is  renormalizable
by renormalizing the coefficients $c_i$. It is clear that by
properly adjusting the additions to the quantities  $c_i$  in
every order of the perturbation theory, i.e., by manipulating the
finite renormalizations, in the limit $\La\to\infty$,  we  can  achieve
the coincidence of the value of every Feynman  diagram  with  its
conventional value (the corresponding scheme is briefly presented
in Appendix~4). This is just what we wanted to prove.

{\bf Acknowledgments.} This work was supported in part (S.~A.~P.)  by
the grant 96-0457 INTAS within the framework of the research
program  of International  Center  of  Fundamental  Physics  in  Moscow
(ICFPM) and the Euler Program of Berlin Free University.

\setcounter{form}{0}
\renewcommand{\theform}{{\rm A}1.\arabic{form}}

\section*{$\protect\vphantom{a}$\hfill Appendix 1}

We compare  the  results  of  calculating  an  arbitrary  Feynman
diagram in the light-front coordinates (with the limit  $\e\to 0$
subsequently taken) and in the Lorentz coordinates. Every diagram
is constructed from summary propagators (\ref{2.6}) and $\Delta^{\ps}$
as  well  as
from the vertices entering interaction Lagrangian (\ref{v.3})  with  the
conditions $|k_-|\ge\e$ and $k_\p^2\ge  v^2$ but  without  the  conditions
$|k_-|\le V$ and $k_\p^2\le V^2$.

We use the formalism presented in \cite{tmf}. Under the condition
$k_\p^2\ge  v^2$, which is equivalent to the  presence  of  the  nonzero
mass in two dimensions, the form of propagator (\ref{2.6}) is admissible
for this formalism. It is easy to see that for all diagrams,  the
index $\om_{\pa}$ of the UV divergence with  respect  to  $k_+$
and  $k_-$  is
negative; therefore, for our theory, there are no  special  cases
described in \cite{tmf}.  The  numerators  of  all  integrands  of  the
Feynman diagrams are polynomials; therefore, we have $\ta>0$ and $\et>0$
(for the notation, see~\cite{tmf}). The basic formula is
 \disn{a.1}{
\s=min(\ta,\om_--\om_+-\m+\et),
\nom}
where the minimum is taken over all subdiagrams, and the required difference between the light-front and
Lorentz-invariant calculations is of the order $\e^\s$.

An external gluon line carrying the label "$+$" contributes $+1$ to the
value of $\om_--\om_+$. A pair of external  fermion  lines  that  are
connected with the continuous fermion line contributes $-1$ (if the
diagram is proportional to $\g^+$ with respect to the labels of  this
pair) or $+1$ (if it, is proportional to $\g^-$) or $0$ (in  the  other
cases) to the value of $\om_--\om_+$. We can see  from  formula (\ref{2.6})
that without considering  the  factors  from  the  vertices,  the
summary gluon propagator of the $\Pi$-line contributes  $-2$  (if  the
propagator carries the labels "$\p\p$")  or $-3$ (if  the  propagator
carries the labels "$\p +$" or "$++$") to the value of $\m$.  The  factors
from the vertices carrying the index "$+$" contribute $+1$ to the value
of $\m$. From formula (\ref{2.5}) with conditions (\ref{v.4a})
taken into  account,
we see  that  the  summary  fermion  propagator  of  the  $\Pi$-line
contributes $-2$ to the value of $\m$. This number increases  if  we
decrease the  number  of  the  additional  Pauli-Villars  fermion
fields.
Analyzing this information, we obtain the  general  form  of  the
diagrams with $\s\le  0$ (in fact, $\s=0$). It is presented in Fig.~1.
\begin{figure}[htb]
\begin{center}
\epsfig{file=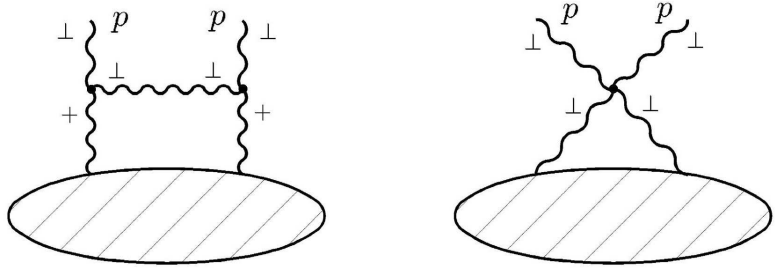,width=11cm}
\caption{
The general form of  the  QCD  diagrams  for  which  there
exists  a  difference  between the light-front and the
Lorentz-invariant calculations; $p$ is an  external  momentum;  the
symbols "$+$" and "$\p$" at the lines denote  the  corresponding  Lorentz
index of the propagator.
}
\end{center}
\end{figure}
The conditions for these diagrams  to  be  trivially  dependent  on
external momenta are fulfilled (see~\cite{tmf});
therefore, for these diagrams, the required difference is
 \disn{a.2}{
Cg_{AB},
\nom}
where $C$ is a constant in the limit $\e\to 0$, i.e., the dependence
on external momenta is absent. There is no  possible  logarithmic
correction $\ln(\e/u)$, because, in view of the  Lorentz  invariance
in the space of $k_+$ and $k_-$, which holds for the Lagrangian, the
quantity $u$ must behave like the "$-$"-component of a vector,  but
there are no such expressions. There could only be the  component
of an external momentum, but, we already know that there  is  no
dependence on it. Therefore, all the difference is compensated by
adding the term of the form
 \disn{a.3}{
A^a_\m g^{\m\n} A^a_\n,
\nom}
to the Lagrangian, i.e., by redefining the coefficient $c_2$
in formula (\ref{v.3}).

\setcounter{form}{0}
\renewcommand{\theform}{{\rm A}2.\arabic{form}}

\section*{$\protect\vphantom{a}$\hfill Appendix 2}

We analyze the possibility of the occurrence of the essential  IR
divergences (i.e., those occurring at any value of  the  external
momenta) in the Feynman  diagrams  in  the  Euclidean  space.  We
consider two cases of the divergences: when integrating over only
the longitudinal momenta (they can appear because of  (the  term
proportional  to  $k_+/k_{\pa}^2$  in  the  gluon  propagator)  and    when
integrating over all momenta (the factors $1/k^2$ in the propagators
also contribute to these divergences).

{\bf First case.}
We study whether a divergence exists if a part (or  all)  of  the
longitudinal loop momenta tend to some finite values.  We  assume
that the external momenta do not take the special  values  (where
the sum of a part of them is equal  to  zero).  In  addition,  we
assume that the transverse loop momenta are such that for all  of
the propagator momenta, we have $Q_\p\ne 0$ (if  this  condition  is
violated, we must take the extra contribution  from $d^2Q_\p$ into
account; this is the second case). Then the divergence can  arise
only if for some gluon line, we have $Q_{\pa}=0$. The factor  in  the
integrand producing the possible divergence has the form
 \disn{b.20}{
\frac{Q_\m n_\n+Q_\n n_\m}{Q_\al n^\al}=-\sqrt{2}
\frac{(Q_\m n_\n+Q_\n n_\m)(iQ_0+Q_3)}{Q_0^2+Q_3^2}
\nom}
and is a pole of an order  not  higher  than  one.  In  the  loop
momentum space, we consider a point where $Q_{\pa}=0$ for a  certain
set of lines. We look for the power $\s$ of the IR divergence. Every
line of this set contributes $-1$ (if it carries the labels $+\p$,
see~(\ref{b.20})) or $0$ (if it carries  the  labels  $\p\p$ or $++$).  We
exclude all lines with the labels $\p\p$ and $++$; then every line of
the set contributes $-1$.  The  differentials  $d^2q_{\pa}$  of  the  loop
momenta  (the  integration  volume  elements)  give  a   positive
contribution to $\s$. We must consider only those loop momenta whose
change (other momenta being fixed) violates the condition $Q_{\pa}=0$
for the lines of the set. The number of such loop momenta is  the
number of lines whose momenta can be taken arbitrarily, i.e., the
number of independent lines. The total positive  contribution  to
$\s$  is  equal  to  twice  this  number.  We  then   find    other
contributions to the IR divergence.

We break all lines of  the  set.  The  diagram  splits  into $n+1$
connected parts ($n=0,1,\dots$).  All  momenta  external  with
respect to the whole diagram enter one part (the external momenta
would otherwise take the  special  values).  We  call  this  part
separated and the other parts nonseparated.

A nonseparated part is a  subdiagram;  if  it  has  the  external
Lorentz labels, then it must be proportional to its external line
momenta carrying these labels. But the factor of  proportionality
cannot contribute to $\s$, because of the Lorentz  invariance,  the
invariance with respect to multiplying $n^\m$ by a complex number,
and the condition $Q_\p\ne  0$. Every external line carrying the label
$k$ ($k=1,2$) gives the factor $k^k$ or $g^{\m k}$;
the line carrying the label "$+$" gives the factor $k^+$, where $k$  is  a
linear combination of the external  momenta  of  the  subdiagram,
i.e., the momenta of the set  lines.  Therefore,  every  external
line carrying the label "$+$" contributes $+1$ to $\s$, and one  carrying
the label $k$ contributes zero. Let $m$ be a summary contribution  to
$\s$ obtained in such a way from the nonseparated parts.

Let $s$ be the number of lines of the set and $r$ be  the  number
of lines of the set external with respect to the separated  part.
Every line carries the label "$+$" at one end; therefore, we have
 \disn{b.21}{
m\ge s-r.
\nom}
The number of the independent lines in the set is equal to $s-n$
(each of the $n+1$ parts gives the $\de$-function, and  one  $\de$-function
is common to the whole diagram). By the definition of $\s$, we have
 \disn{b.22}{
\s=2(s-n)+m-s.
\nom}
Using (\ref{b.21}), we obtain
 \disn{b.23}{
\s\ge 2(s-n)-r.
\nom}
Every nonseparated part has a minimum of two external lines. This
means that all parts (separated and nonseparated) together have a
minimum of $2n+r$ external lines of the set. All  these  lines  are
pairwise connected with each other. Consequently, we have
 \disn{b.24}{
s\ge\frac{1}{2}(2n+r)=n+\frac{r}{2},
\nom}
whence we obtain $\s\ge 0$, i.e., the  divergence  cannot  be  worse
than logarithmic.

We find the general case producing this divergence. In this case,
all the above-cited inequalities must reduce to equalities, i.e.,
we have $m=s-r$, and all the nonseparated parts  are  two-point
diagrams. The general form of such a diagram is shown in Fig.~2.
\begin{figure}[htb]
\begin{center}
\epsfig{file=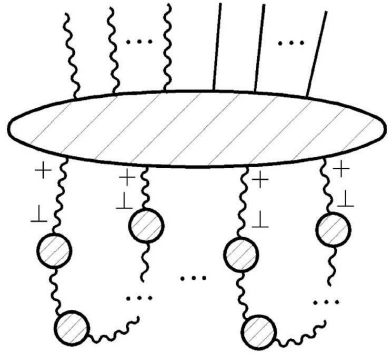,width=6cm}
\caption{
The general form of the diagram producing the  essential
IR divergence in QCD; the symbols "$+$" and "$\p$" by the lines mean that
the propagator has the corresponding Lorentz index.
}
\end{center}
\end{figure}
In this diagram every chain giving essential  IR
divergences contains the factor
 \disn{b.28}{
\De_{+\rho} G^{\rho\n}\De_{\n\al},
\nom}
where $G^{\rho\n}$ is the nearest to the separated part nonseparated
one which is one-particle irreducible subdiagram and  $\De$ are the
propagators of the lines entering to the $G^{\rho\n}$.
The divergence takes place only when the index  $\rho$ in
(\ref{b.28}) takes values $1,2$.

{\bf Second case.} We now assume  that  a  part  or  all  of  the
four-dimensional momenta  lend  to  certain  finite  values.  The
consideration is similar. The gluon propagator now gives  a  pole
of the second order, whereas  the  fermion  propagator  gives  no
poles, because of the presence  of  the  mass.
Notice, that at the investigation of IR divergences one can replace
all fermion lines by $1/M$ which makes impossible getting $1/M$ to
the numerator. Besides in this case one can replace $R$  by
1 and $\hat\mu^2$ by $\mu^2$ in the propagator of gluon which excludes
$\La$ from the consideration. That is why from  dimensional
considerations,  it  follows  that  every    nonseparated    part
contributes no less than four minus the number  of  its  external
lines to $\s$. This is also true for the divergent diagrams with the
stipulation that for the two-point diagram, we must  subtract  the
part  independent  of  the  momentum  and  proportional  to  $g^{\m\n}$
together with the divergent part, i.e.,  we  must  normalize  the
gluon mass to zero in every order. The latter  does  not  violate
the Ward identities in the limit of removing the  regularization.
Therefore, we now have
 \disn{b.29}{
m\ge 4n-(2s-r),\no
\s=4(s-n)+m-2s,
\nom}
whence we obtain
 \disn{b.32}{
\s\ge r>0,
\nom}
i.e., there is no divergence.

\setcounter{form}{0}
\renewcommand{\theform}{{\rm A}3.\arabic{form}}

\section*{$\protect\vphantom{a}$\hfill Appendix 3}

In the framework of the perturbation theory constructed based  on
Lagrangian (\ref{v.1})-(\ref{v.3}),
we consider an arbitrary diagram  with  three
gluon external lines and investigate its structure  with  respect
to the labels of the global gauge group. This structure  has  the
form $A^{abc}$. The gluon propagator is  proportional  to  $\de^{ab}$,  the
three-point  gluon  vertex  is  proportional  to  $f^{abc}$,  and    the
four-point vertex is proportional
to  $f^{abe}f^{cde}$  or  $\de^{ab}\de^{cd}$.  In
addition, there can be fermion loops.

Every fermion loop gives the factor
 \disn{c.1a}{
{\rm Re Tr}(\la^a\dots\la^c),
\nom}
if the number of vertices in the loop is even or
 \disn{c.1b}{
i{\rm Im Tr}(\la^a\dots\la^c),
\nom}
if this number is odd (similar to the Furry  theorem  in  quantum
electrodynamics). We consider only the factors  related  to  the
gauge labels. From (\ref{v.7}), it follows that
 \disn{c.3}{
\la^a_{\al\be}\la^a_{\g\de}=2\ls \de_{\al\de}\de_{\be\g}-
\frac{1}{3} \de_{\al\be}\de_{\g\de}\rs.
\nom}
We can also write
 \disn{c.4}{
{\rm Tr}\ls\la^a\la^b\la^c\rs=i2f^{abc}+2d^{abc},
\nom}
whence it follows that
 \disn{c.5}{
f^{abc}=-\frac{i}{4}\ls{\rm Tr}\ls\la^a\la^b\la^c\rs-
{\rm Tr}\ls\la^c\la^b\la^a\rs\rs.
\nom}

We replace all $f^{abc}$ in the diagram with the right-hand side  of
formula (\ref{c.5}) and then replace all resulting expressions of the
form $\la^a\la^a$ according to formula (\ref{c.3}).
As a result, only  three
matrices $\la$ carrying three external labels of the diagram  remain.
These matrices are connected with each other by their  labels  of
the fundamental representation (because after formula (\ref{c.3})  is
used, only the Kronecker symbols remain),  i.e., $A^{abc}$ consists
of terms of the form
 \disn{c.6a}{
{\rm Tr}(\la^a\la^b\la^c).
\nom}

After using formula (\ref{c.5}), we have the factor $i^N$,  where  $N$  is
the number of three-point gluon  vertices.  The  use  of  formula
(\ref{c.3}) gives no new imaginary factors.  The  initial  expression
consists of the real quantities $\de^{ab}$ and  $f^{abc}$
and  the  quantities
(\ref{c.1a}) and (\ref{c.1b}), of which the first is real and the second is
imaginary; therefore, it is proportional to $i^{N'}$, where $N'$ is the
number of fermion loops with an odd number  of  vertices.  It  is
easy to show that for  the  diagram  with  three  gluon  external
lines,  the  quantities  $N$  and  $N'$  have  different    parities.
Therefore, for the powers of $i$ to be consistent, it is  necessary
that the sum of the expressions of  form  (\ref{c.6a})  be  imaginary,
whence, using (\ref{c.4}), we conclude that
 \disn{c.7}{
A^{abc}\sim f^{abs}.
\nom}

We now similarly consider the structure of an  arbitrary  diagram
with four gluon external lines. It has the form $A^{abcd}$. The  quantity
$A^{abcd}$ consists of terms of the forms
${\rm  Tr}(\la^a\la^b\la^c\la^d)$ and
${\rm  Tr}(\la^a\la^b){\rm  Tr}(\la^c\la^d)$
(the latter expression is explicitly real). It is  easy  to  show
that  for  the  diagram  with  four  gluon  external  lines,  the
quantities $N$ and $N'$ have the same parities. Therefore, the sum of
the indicated expressions, which compose $A^{abcd}$, must  be  real.
We  can  represent  the  quantity  ${\rm Tr}(\la^a\la^b\la^c\la^d)$)
as  a  sum  of  its
symmetrized part (explicitly real) and expressions  of  the  form
$if^{abe}{\rm  Tr}(\la^c\la^d\la^e)$.
In view of the reality condition,  the  latter
expression must be a sum of quantities  of  the  form  $f^{abe}f^{cde}$.
Direct  calculation  shows  that  the  symmetrized  part  of  the
quantity ${\rm Tr}(\la^a\la^b\la^c\la^d)$
is proportional to the symmetrized part  of  the
quantity  $\de^{ab}\de^{cd}$.  We  can  therefore  conclude  that
$A^{abcd}$ consists of terms of the forms
 \disn{c.9}{
f^{abc}f^{abc} \quad {\rm and} \quad \de^{ab}\de^{cd}.
\nom}

\setcounter{form}{0}
\renewcommand{\theform}{{\rm A}4.\arabic{form}}

\section*{$\protect\vphantom{a}$\hfill Appendix 4}
We shall find a form of
the values which are necessary to add to Feynman diagrams
in order to in the limit $\La\to\infty$ make them
finite and coincident
with its value calculated using dimensional regularization.
We consider a diagram in Euclidean space constructed from
propagator (\ref{isp6}), from  the  propagator $\Delta^{\ps}$,  and  from  the  vertices
entering (\ref{2.2}) with restrictions (\ref{v.6}) removed.
The other vertices
entering (\ref{v.3}) are involved only in reducing the subdiagrams of the
preceding orders of the  perturbation  theory  to  the  "correct"
(i.e.,  calculated  using  dimensional   regularization)    value
including cancellation of divergences.

It is clear that the summary fermion propagator can be represented as
 \disn{d.1}{
\Delta^{\ps}=\frac{k_\m\g^\m}{k^2-M_0^2+i0}R'+\frac{M_0}{k^2-M_0^2+i0}R'',
\nom}
where $R'$ and $R''$ are cutoff factors  that  properly  decrease
and $R',  R''\str{\La  \to  \infty}  1$. We represent the diagram as
a sum such that  one  summand  contains  only  one  term  of  the
numerator of every fermion propagator. Each of these summands can
be represented as
 \disn{d.2}{
I=\int dk F(k,p)f_{\La}(k,p),
\nom}
where $dk$ represents all volume elements, $f_{\La}(k,p)$ is the product
of all factors $R'$ and $R''$ entering tlie fermion propagators and of
all factors $R$ entering boson  propagators  (\ref{2.6}),  $k$  denotes  the
integration momenta, and $p$ denotes the external  momenta  of  the
diagram.

It is  evident  that  we  can  find  a  function $\hat F(k,p)$ that  is
polynomial in $p$, is Lorentz invariant, and has  no  nonintegrable
IR singularities such that for $\hat  F(k,p)$ and $F(k,p)$,  a  number  of
the first terms of their asymptotic expansions at  infinity  with
respect to $k$ coincide and the  difference  $F(k,p)-\hat  F(k,p)$  is
integrable (see the refinement below). Therefore, we can write
 \disn{d.3}{
I=\int dk (F(k,p)-\hat F(k,p))f_{\La}(k,p)+
\int dk \hat F(k,p)f_{\La}(k,p).
\nom}
Up to corrections of the order $1/\La$, we can neglect the factor
$f_{\La}(k,p)$ in the first integral (because  the  integral  converges
even in the absence of this factor). We can then assume that this
integral is calculated using dimensional  regularization  and  we
can split it into two integrals  (assuming  that  they  are  both
renormalized by dimensional  regularization).  As  a  result,  we
obtain the expression
 \disn{d.4}{
I=\int^{dim} dk F(k,p)-\int^{dim} dk \hat F(k,p)+
\int dk \hat F(k,p)f_{\La}(k,p).
\nom}
Using the expansion for the function $f_{\La}(k,p)$ in the last integral
in (\ref{d.4}) with respect to $p$ in the neighborhood  of  the  origin
(this expansion is well defined for any $k$), we can now show that
the  last  integral  in  (\ref{d.4})  is  a  polynomial  in  $p$   plus
corrections of the order $1/\La$.

We must make the following refinement  of  these  considerations.
Before subtracting the overall divergence of the diagram, we must
verify that the diagram has no subdivergences with respect  to  a
part of the integration variables.  Such  subdivergences  can  be
produced by subdiagrams (which  is  taken  into  account  by  the
renormalization in the lower orders)  or  by  the  divergence  on
integrating over a part of the momentum components. For  the  QCD
in the gauge $A_-=0$,  the  latter  is  possible  because  of  the
improper decrease of the propagator in the direction  $k_\p$.  It  is
known \cite{wein} that this divergence is present if the index of the UV
divergence  with  respect  to  a  part  of  the  components    is
nonnegative. From the structure of the propagators,
we can see that in the Euclidean space, the  subtraction  of  the
overall divergence cannot decrease only the index $\om_\p$ of the UV
divergence with respect to $k_\p$. This results in the necessity  to
preliminarily subtract, the subdivergence with respect to $k_\p$, the
subdivergence being in general dependent on the projections $p^\n n_\m$ of
the external momenta in an arbitrary nonpolynomial way.  Analysis
of the diagrams for the Green's functions shows that we have
$\om_\p\ge  0$ for only the one-loop two-point diagrams and that
$\om_\p=0$.
These diagrams have only one external momentum.  Therefore,  they
cannot have a nonpolynomial dependence on $p^\n n_\n$, because  of  the
invariance with  respect  to  multiplying  the  vector $n_\n$ by  a
complex number. One must take into account  that,
for the  diagrams  of  the  Green's
functions, the vector $n_\n$ with the nonconvolute  index  $\n$  cannot
stand in the numerator, because this gives  zero  on  convolution
with the propagators. If we consider the one-particle irreducible
vertex parts, whose diagrams do not satisfy the  last  condition,
the divergent parts can be nonpolynomial \cite{bas1,bas2}, and the number
of diagrams that are divergent with respect to $k_\p$ can  be  much
larger.

It seems that if the integral of the form, described at the beginning
of this Appendix, converges then the result of it' s calculation cannot
depend on the vector $n^*_\n$ (complex conjugated to
$n_\n$ up to a factor). But this would be true only if this
integral be an analitical function of the
complex vector $n_\n$.  However the derivative of our integral
with respect to complex vector $n_\n$ can be nonconvergent
due to the rising of infrared singularity (of the power
of the pole in $(nk)$). Therefore the result of the calculation
of  the diagram, and, hence, it's divergent part can depend
on $n_\n$ and on $n^*_\n$. It is seen from the equation (\ref{isp6})
for the propagator that the integrands
we considered  are invariant  with respect to a  multiplication
of the vector $n_\n$ by a complex number (and
of the $n^*_\n$ by complex conjugate number).
This allows to conclude that up to corrections of the  order
$1/\La$  the  difference  between  the  diagram  described  in   the
beginning of this appendix (more precisely,  the  finite  sum  of
such diagrams of the given order) and its value calculated  using
dimensional regularization (i.e., the similar sum  of  the  first
integrals in (\ref{d.4})) reduces to a  polynomial  in  the  external
momenta with  coefficients containing factors $N_{\al\be}$
(see the definition of this value in formula (\ref{isp7})).

\end{document}